\documentclass[conference]{IEEEtran}
\IEEEoverridecommandlockouts
\usepackage{amsmath,amssymb,amsfonts}
\usepackage{algorithmic}
\usepackage{graphicx}
\usepackage{textcomp}
\usepackage{xcolor}
\usepackage{siunitx}
\usepackage{algorithmic} 
\usepackage{multirow}
\usepackage{wrapfig}
\usepackage{biblatex}
\usepackage{tikz}
\usepackage{lipsum}

\newcommand\copyrighttext{%
  \footnotesize \textcopyright 2019 IEEE. Personal use of this material is permitted. Permission from IEEE must be obtained for all other uses, in any current or future media, including reprinting/republishing this material for advertising or promotional purposes, creating new collective works, for resale or redistribution to servers or lists, or reuse of any copyrighted component of this work in other works.}
\newcommand\copyrightnotice{%
\begin{tikzpicture}[remember picture,overlay]
\node[anchor=south,yshift=30pt] at (current page.south) {\fbox{\parbox{\dimexpr\textwidth-\fboxsep-\fboxrule\relax}{\copyrighttext}}};
\end{tikzpicture}%
}
\def\BibTeX{{\rm B\kern-.05em{\sc i\kern-.025em b}\kern-.08em
    T\kern-.1667em\lower.7ex\hbox{E}\kern-.125emX}}

\newcolumntype{P}[1]{>{\centering\arraybackslash}p{#1}}

\begin{document}

\title{Reinforcement Learning based Orchestration for Elastic Services\\
}

\author{\IEEEauthorblockN{Mauricio {Fadel Argerich}, Bin Cheng, Jonathan F{\"u}rst}
\IEEEauthorblockA{\textit{NEC Laboratories Europe, Heidelberg, Germany}\\
mauricio.fadel@neclab.eu, bin.cheng@neclab.eu, jonathan.fuerst@neclab.eu}}

\maketitle

\begin{abstract}
Due to the highly variable execution context in which edge services run, adapting their behavior to the execution context is crucial to comply with their requirements.
However, adapting service behavior is a challenging task
because it is hard to anticipate the execution contexts in which it will be
deployed, as well as assessing the impact that each behavior change will
produce. In order to provide this adaptation efficiently, we propose a Reinforcement Learning (RL) based Orchestration for Elastic Services. We implement and evaluate this approach by adapting
an elastic service in different simulated execution contexts and comparing its
performance to a Heuristics based approach. We show that elastic services achieve high precision and requirement satisfaction rates while creating an overhead of less than 0.5\% to the overall service. In particular, the RL approach proves to be more efficient than its rule-based counterpart; yielding a 10 to 25\% higher precision while being 25\% less computationally expensive. 
\end{abstract}


\copyrightnotice

\vspace{-1.9em}

\section{Introduction}
Increased data traffic and network utilization are one of the biggest
challenges for network operators nowadays. One of the reasons is the massive
amount of data generated by devices in the edge in the context of the Internet
of Things (IoT). Edge computing~\cite{garcia2015edge} allows network operators
to reduce network stress and improve service responsiveness by allocating
computation closer to data producers and consumers. Nonetheless, edge
processing hardware is constrained and heterogeneous, which makes it hard to
provide cloud-like elasticity features (i.e., scale out) that are necessary to
react to the burstiness of typical IoT loads (e.g., loads that are based on
user behavior or interaction). For example, the load of a local edge server
that serves an augmented reality (AR) application~\cite{chen2018snaplink}
is directly correlated to the number of active users. Too many active
users result in exhaustive response times and poor user experience.

To address such problems, we proposed in our previous
work~\cite{fuerst2018elastic}, edge hosted services need to dynamically
adapt to the current execution context to better comply with
their non-functional application requirements or Service Level Objectives
(SLOs) together with the execution framework. We call this concept ``Elastic
Services''.
However, adapting service behavior to a given context is a challenging
task, because it is hard to anticipate the scenarios the software might
encounter (e.g., different loads or wireless link
quality~\cite{baccour2012radio}) as well as to assess the impact
that each behavior change will produce.

In this work, we propose a Reinforcement Learning (RL) based orchestration to adapt services and applications behavior during runtime so
they best adhere to non-functional requirements like response time. Our
approach starts exploring different behavior alternatives, and learns --based
on its own experience-- the best behavior for the current and potentially
complex execution context. We implement and evaluate a prototype of this approach that provides high satisfaction of service requirements
while being computationally inexpensive.

Our main contributions are as follows:

\begin{itemize}
    \item Definition of a RL based approach and its elements---actions, states
        and reward---to adapt services to their current
        execution context.
    \item Evaluation of the RL based approach and its comparison to a Heuristics based
        approach, by means of simulation.
\end{itemize}

\section{Motivation}%
\label{sec:motivation}

To illustrate the complexity of dynamically deciding on the best adaptation, we
introduce a typical video analysis application: the Lost Child Application. The
application works in the following way: if a child goes missing, law
enforcement asks their parents for photographs of the child and a facial
recognition classifier is trained with them. Then, a service is deployed using
existing connected cameras and edge servers in the city to analyze the video
feeds and to locate the child. When a matching face is found, a notification is
sent to nearby law enforcement officers. This application can be split in two
components: (1) an offline module, which is trained with pictures of the child
in a server and (2) an online module, a face detection and matching service that is deployed in several devices and is in charge of finding the child. We will focus our attention on the latter.

The face detection and matching service is composed by the following steps, translated as functions
of the service:
\begin{enumerate}
  \item \textbf{Capture image}: the image is captured by the camera.
  \item \textbf{Image preprocessing}: different preprocessing steps such as
      resizing, colorization, etc., are performed to improve the accuracy of
      the face detection and recognition.
  \item \textbf{Face detection}: a face detection classifier is applied to the
      preprocessed image and each face is extracted and returned.
  \item \textbf{Face Recognition}: for each face that was found in the image,
      the previously trained classifier that matches the face of the lost child
      is applied.
  \item \textbf{Notify law enforcement}: a notification is sent to nearby
      police officers indicating the location of the child.
\end{enumerate}

A small end-to-end latency is necessary to ensure a high frame sampling rate,
so the event of missing the child is unlikely to happen, even if it only
appears briefly on the video feed, e.g., when the child is moving. The requirement is to have at least one frame analyzed per second, which
is translated into having a maximum end-to-end latency of \SI{1}{s} for each
analyzed frame. At the same time, it is desired that the service performs with
the highest precision possible.

In order to comply with these requirements, it is necessary to adapt the
service behavior to the current execution context:

\begin{itemize}
    \item The hardware capabilities of the device in which the application or
        service executes affects its performance, e.g., a device that can use a
        graphic accelerator to speed up matrix multiplication will be able to
        process faster neural network algorithms than one without.
    \item Different camera input results in different performance. For example,
        in the face detection and matching service, analyzing a frame with 5 persons is less
        computationally expensive than analyzing a frame with 100 persons. All
        of these factors generate different latencies for processing images and
        therefore affect the requirements satisfaction. 
\end{itemize}

\vspace{-0.5em}
\subsection{Adaptation Knobs}%
\label{sub:adaptation_knobs}

For each function of the service, a number of parameters can be modified in order to modify the service's performance.
The optimal parameters configuration varies according to the specific
requirements of the service, its execution context and its inputs.
This means that these values need to be changed for every device and also
during runtime, in order to obtain the best possible performance.

\begin{figure}[ht]
\vspace{-1em}
\setlength\abovecaptionskip{-0.2\baselineskip}
    \centering
    \includegraphics[width=\linewidth]{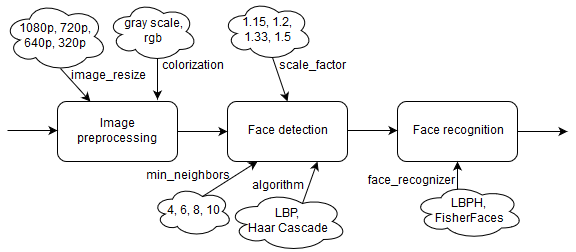}
    \caption{Different behavior adaptations for the face detection and matching service}
    \label{fig:lost-child-behavior}
\end{figure}

Furthermore, the number of alternative behaviors grows exponentially with the
number of parameters and the values each parameter can assume. As shown in
Figure~\ref{fig:lost-child-behavior}, even for a simple service with three
different functions many parameters can be adapted.
There are six adaptable parameters, three of them have four different values
(image\_resize, scale\_factor and min\_neighbors), and three of them have two
different values (colorization, face detection algorithm and face\_recognizer).
This means that there are $4^3 \cdot 2^3 = 512$ different configurations that
can be used to adapt the service's behavior to a given execution context.

Because of the high number of configurations together with the uncertainty of their
impact on service requirements, finding the best configuration of parameter
values for a given execution context is a complex task. Choosing this
configuration manually is also ineffective, as there is no universal set of
values that works properly across all devices and contexts.
An automatic approach is needed, one that is able to learn from the
service, its inputs and its execution context, to decide what is the best
parameters configuration for the current execution context. 

\vspace{-0.5em}
\section{Elastic Services}

In this section we introduce the programming model for service developers to
easily define elastic services and also the underlying edge computing
framework to support such a programming model.  

\subsection{Programming Model of Elastic Services}

To simplify the development of elastic services, we extend the traditional
dataflow-based programming model~\cite{akidau2015dataflow} to support service
elasticity in the following way. 

First, service developers break down the logic of their services into small
processing functions. Each processing function is called an \textbf{operator}.
However, different from the traditional dataflow programming model, these
operators are parameterized to change their internal execution during runtime,
meaning that each operator is associated with a set of parameters, and by
changing these parameters, we can control the behavior of the operator on the
fly. For example, a clustering operator can use a parameter to control which
clustering algorithm should be applied for this operator at runtime. The
implementation of an operator can be mapped to various of dockerized
application images that are deployable and executable in any docker-based
environment, either in the cloud or at edges.  

Once operators are defined and their implementation images are provided,
service developers can start to specify a \textbf{service topology} to
represent the abstract processing logic of their service. A service topology
consists of a set of linked tasks and each task is an annotated operator with a
specific granularity. The granularity of a task determines how many task
instances of the same operator should be instantiated at runtime, based on the
available data.

A service topology must be triggered explicitly by a user-definable
\textbf{requirement}, issued by a consumer or any application on-demand. The
requirement defines when and how the defined service topology should be
instantiated, which is the main challenge to be addressed by service
orchestration in general. One important part of the requirement is to cover the
QoS defined by users in terms of required latency, reduction of bandwidth
consumption, or any other high level metrics. The goal of our service
orchestration is to achieve and ensure the required QoS continuously by making
orchestration decisions adapted to the ongoing workload and also any
environment changes. 

\subsection{FogFlow: Edge Computing Framework for Elastic Services}

FogFlow\cite{cheng2018fogflow} is a distributed execution framework that dynamically orchestrates
elastic services over cloud and edges, in order to reduce internal bandwidth
consumption and offer low latency. The unique feature of FogFlow is
context-driven, meaning that FogFlow is able to orchestrate dynamic data
processing flows over cloud and edges based on three types of contexts,
including:

\textbf{System context}: available resources which are changing over time. The
resources in a cloud-edge environment are geo-distributed in nature and they
are dynamically changing over time; As compared to cloud computing, resources
in such a cloud-edge environment are more heterogeneous and dynamic.

\textbf{Data context}: the structure and registered metadata of available data,
including both raw sensor data and intermediate data. Based on the standardized
and unified data model and communication interface, namely NGSI, FogFlow is
able to see the content of all data generated by sensors and data processing
tasks in the system, such as data type, attributes, registered metadata,
relations, and geo-locations. 

\textbf{Usage context}: high level intents defined by all different types of
users (developers, service consumers, data providers) to specify what they want
to achieve. For example, for service consumers, they can specify which type of
results is expected under which type of QoS within which geo-scope; for data
providers, they can specify how their data should be utilized and by whom. 

\begin{figure}[ht]
\vspace{-1.2em}
\setlength\abovecaptionskip{-0.4\baselineskip}
    \centering
    \includegraphics[width=1.0\linewidth]{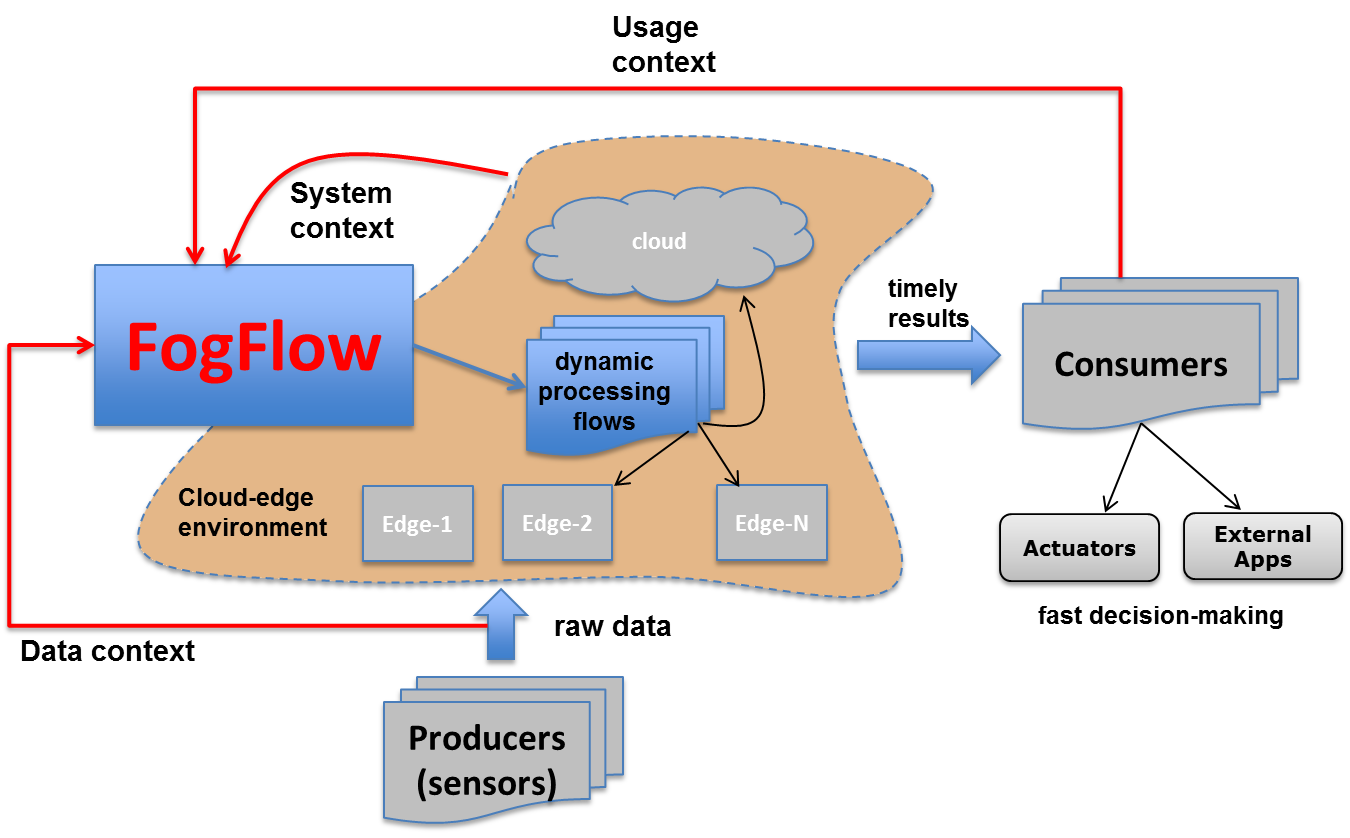}
    \caption{FogFlow System Overview}
    \label{fig:figures/fogflow_architecture}
\end{figure}

As shown in Figure~\ref{fig:figures/fogflow_architecture}, by leveraging these
three kinds of context, FogFlow is able to orchestrate elastic IoT services in
a more intelligent and automatic manner. The overall design of FogFlow has been
presented in our previous paper~\cite{cheng2018fogflow}. In this paper we focus on
the algorithms of service orchestration, which can be applied by the FogFlow
system framework to support elastic services. 

\section{Dynamic Orchestration}
In order to generalize our approach over services with different numbers and
types of requirements, we model the problem as a constrained optimization
problem. Specifically,(1) we model requirements as constraints, e.g., to
process documents with an end-to-end latency less or equal than \SI{1}{s} or to
run at a cost of less or equal than \$10 per hour and (2) we model
service performance, such as precision, accuracy or battery consumption, as
objective.
There is an important difference between the objective and the constraints:
whereas the constraints define a maximum or minimum value for the variable
involved (e.g., latency, cost, etc.), the objective does not have a minimum or
maximum value expected.
In this way, we can define the service requirements as:

\begin{equation*}
\label{eq:opt-problem}
\begin{aligned}
& \underset{\theta}{\text{maximize}}
& O(\theta) & & \\
& \text{subject to}
& c_i(\theta) \leq C_i, & \; i = 1,\ldots,N
\end{aligned}
\end{equation*}

where:
\begin{itemize}
    \item $\theta$: is the configuration of parameters used for all of the
        operators
    \item $O(\theta)$: represents the objective of the service, which is
        determined by the configuration of parameters used
    \item $c_i(\theta)$: is a constraint to the service (such as latency), also
        determined by $\theta$
    \item $C_i$: is the constraint target (e.g., \SI{1}{s})
    \item $N$: is the total number of constraints.
\end{itemize}

The developer is in charge of defining the service requirements along with the
metrics to monitor them, as well as the parameters that can be adapted and the
values they can assume. During runtime, the system is in charge of finding the
best configuration of parameter values that maximize (or minimize) the
objective while respecting the constraints.

To ensure the correct functioning of the service, the dynamic orchestration has
two main requirements:

\begin{itemize}
    \item \textbf{Rapid response.} It must adjust the service behavior rapidly
        to keep up with the context changes during runtime. A slow response
        might mean the violation of the requirements if the resources are
        further constrained, or the loss of improvement in the objective if
        more resources become available.

    \item \textbf{Low overhead.} The dynamic orchestration must not create a considerable overhead for the system, so most of the execution time is still
        used for the service.
\end{itemize}

We present two approaches that implement this constraint optimization for
service orchestration: (1) a heuristic based approach and (2) a reinforcement learning based
approach.

\subsection{Heuristics based Orchestration}
First, we develop a heuristic that is based on the assumption of a linear trade-off
between the objective of the service and its constraints. This linear trade-off
is often seen in algorithms: more computing intensive implementations are slow,
but produce more accurate results, while less computing intensive algorithms
are faster but result in less accurate results. Note that even though this
trade-off is present in the use case of the face detection and matching service for the Lost Child
application (see Section~\ref{sec:motivation})
this is an assumption which highly depends on the service implementation and
the chosen algorithms. Current online streaming services use a similar adaptive logic, but in these cases this adaptive behavior must be specifically implemented for the service, whereas in our approach the service developer is abstracted from its implementation.

\begin{figure}[ht]
\setlength\abovecaptionskip{-0.4\baselineskip}
\centering
\includegraphics[width=\linewidth]{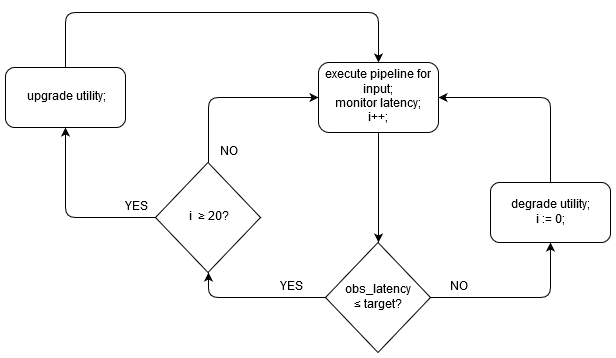}
\caption{Control flow diagram of the heuristics based orchestration}\label{fig:adaptive-heuristic}
\vspace{-1.5em}
\end{figure}

Our heuristic works as follows: to begin, all of the different possible
configurations of parameter values are constructed and sorted by their
objective value (e.g., expected accuracy).
The configuration with the highest objective value is used to process the first
input. Service requirements are monitored and if they are not satisfied, the
performance is degraded by using the immediate lower configuration, which is
expected to improve the requirements satisfaction. If the requirements are
met for a number of continuous steps, then the performance is upgraded.
If this upgrade still satisfies the requirements, then the process is repeated
until requirements are not satisfied anymore, and then the performance is
degraded by utilizing the last configuration that worked within the
requirements. Figure~\ref{fig:adaptive-heuristic} shows a control flow diagram
for this approach.
\vspace{-1em}

\subsection{RL based Orchestration}
Second, we develop a Reinforcement Learning (RL) based optimization to find the
best configuration of parameter values during runtime. The RL based orchestration does not require a linear trade-off between service objectives and constraints. Instead, it learns through its own experience. This makes the RL based orchestration more
flexible and able to generalize over more services than the implemented
heuristic.

In RL, an agent can sense its environment and take actions that affect the
state of the environment and generate a numerical reward. The agent
does not know in advance what actions should be taken in each state, its
objective is to find the action that will create the highest reward for each
different state~\cite{sutton1998reinforcement}. 

In our RL setup the agent represents the service, and its environment can be
seen as its execution context. The agent can take actions to adapt to different
states of its context, in order to achieve its goal of performing with the
highest possible performance while respecting the given constraints. Also, we
frame our problem in a discrete time setting, in which each time step
corresponds to the full processing of an input.

In order for the RL approach to be computationally inexpensive, we use tabular
Q-learning~\cite{watkins1989learning}. Tabular Q-learning defines what action
should be taken in each state by maintaining a table in which there is one row
for each state, and one column for each action. The value in each cell is the
expected reward of taking a specific action while being in a given state.
Because of this, it is of particular interest the definition of the actions and
states. In our case, we define them as:
\begin{itemize}
    \item Actions: each different configuration of parameter values
    \item States: the current execution environment's status, the requirements'
        satisfaction in last step and the last configuration of parameter
        values used
\end{itemize}

When the service has just started, different
parameter configurations are chosen at random and profiled in an online manner.
After this process has been performed a number of times, the agent knows which configurations perform better in each state by using the $Q$ table.

More specifically, two different states-actions configurations are defined and
implemented:

\textbf{Configuration 1}
\begin{itemize}
\setlength\itemsep{0em}
    \item States: Last latency as \% of target [3 values (0-80, 80-100,
        100-$\infty$)], number of last configuration used
    \item Actions: Number of configuration to be used
\end{itemize}

\textbf{Configuration 2}
\begin{itemize}
\setlength\itemsep{0em}
    \item States: Last latency as \% of target [3 values (0-80, 80-100,
        100-$\infty$)], current CPU availability \% [3 values (0-50, 50-80,
        80-100)], number of last configuration used
    \item Actions: Number of configuration to be used
\end{itemize}

\begin{figure}[ht]
\vspace{-1em}
\setlength\abovecaptionskip{-0.1\baselineskip}
\centering
\includegraphics[width=\linewidth]{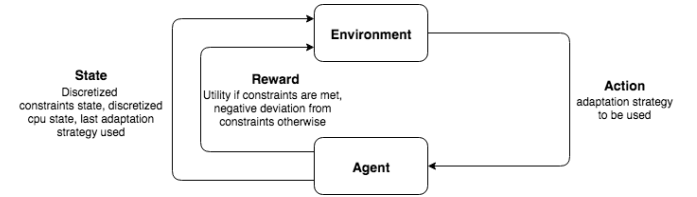}
\caption{Diagram of the RL based orchestration, configuration 2}\label{fig:adaptive-rl_lat_16actions}
\vspace{-1em}
\end{figure}

The reward indicates how well the agent is performing according to its
objective. In our case, the reward is defined as the objective (measured in a
metric selected by the developer e.g. precision) if requirements are satisfied,
or the negative deviation of its performance according to the requirements if
these are not fulfilled. Mathematically, we define the reward function
for taking an action $a$ in a given time step $t$ as:
\begin{equation}
\label{eq:rl-reward}
R_{t,a} =
\begin{cases}
    O_{t-1}, & \text{if}\ \forall c_i \leq C_i \\
    -\sum_{i=0}^{N} \frac{c_{i,t-1}}{C_i} \  \forall c_{i,t-1} > C_i, & \text{otherwise}
\end{cases}
\end{equation}

\section{Evaluation}
\subsection{Simulator}

To test the orchestration approaches, we have implemented a simulator in which
developers can define elastic services, by defining a pipeline of operators,
and the adaptable parameters for each operator. Once a service has been
defined, its functioning can be profiled on a set of inputs and devices by
using a profiler which tests all of the different parameters configurations
repeatedly and record the metrics that will be later used by the simulator.

We implement the face detection and matching service for the Lost Child application in this framework and profile it on a
Raspberry Pi 3B+ device with different inputs. The inputs used are collages
with different numbers of faces, from 6 to 192, taken from the dataset
Faces94~\cite{spacek2007collection}. 

\subsection{Environments and Datasets}
To define the RL framework, we specify different environments with the
states-actions configurations. The implemented interface has been inspired by
OpenAI Gym environments~\cite{brockman2016openai}.

We simulate the service execution on a shared device, which means that the CPU
availability varies over time due to the other services concurrently running.
In every step, the environment simulates the CPU availability as a Markov Chain
varying from 0.3 to 1.0. In this chain, the CPU availability for a new step has
a probability of changing with respect to the previous step of 0.1, and the
variation will be randomly drawn from a $N(0.1,0.1)$ distribution.  

We create three different datasets by combining the inputs previously defined:
\begin{enumerate}
    \item Fixed input: $1000$ frames with 48 faces each.
    \item Variable input: $1100$ frames with varying number of faces in each
        image, from 6 faces to 192. The dataset is composed by blocks of 100
        continuous frames with the same number of faces. The blocks are
        arranged in the following way: 6, 12, 24, 48, 96, 192, 96, 48, 24, 12,
        6; where each number represents the amount of faces in the block of 100
        frames.
    \item Full day input: a dataset that simulates the whole input of a full
        day at a train station was built. There are 86400 images, one for every
        second of the day. The inputs from the previous item were used,
        but the number of faces varies over time to simulate peak and low traffic hours.
\end{enumerate}

In addition, we implement another version of the RL environment. This version
models the random CPU availability as the previous version but incorporates
random inputs. The inputs have different number of faces and are chosen at
random from the inputs used by the variable input dataset. Because the number
of faces in each input does not change every second in reality, in each step
there is a 0.1 probability of changing the input with a random sample and 0.9
probability of keeping the same input as in the previous step.

\subsection{Results}
We simulate the face detection and matching service 50 times with each dataset and orchestration approach. In the case of RL based approaches, the Q-table is started with 0 values for the first simulation and the following ones use a prepopulated Q-table using the values produced by the previous simulations.  Table~\ref{tab:results} depicts the results. 

Firstly, we implement and test a static service with two different configurations: one optimized for a high precision service, and another one optimized for a high latency satisfaction. We can see that the high precision service has a precision of \SI{1}{} --this impressive high value can be achieved thanks to the simple dataset used-- but fails to satisfy the latency requirement very often. The high latency satisfaction service does the opposite, complies with the latency requirement in most cases but offers a very low precision.

Secondly, we implement and test an elastic service using the Heuristics based dynamic orchestration. We can see how the elasticity of the service helps it to get a good trade-off between precision and latency satisfaction rates. 

\begin{figure}[ht]
\setlength\abovecaptionskip{-0.4\baselineskip}
\centering
\includegraphics[width=\linewidth]{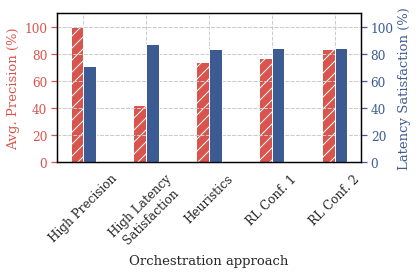}
\caption{Average precision and latency satisfaction over all datasets for the different orchestration approaches.}\label{fig:results-summary}
\vspace{-1.5em}
\end{figure}

Finally, we also test an elastic service using both configurations of RL for its dynamic orchestration. Both RL configurations offer a high latency satisfaction as the Heuristics approach while improving the application's precision
by a margin of 10--25\%, as shown on figure \ref{fig:results-summary}. Moreover, the RL configuration 2 -- which uses the last
requirement's satisfaction, last configuration used and current CPU
availability---shows a better performance than the other configuration; because the agent can take better decisions with more information.

\begin{figure}[ht]
\vspace{-1em}
\setlength\abovecaptionskip{-0.4\baselineskip}
\centering
\includegraphics[width=\linewidth]{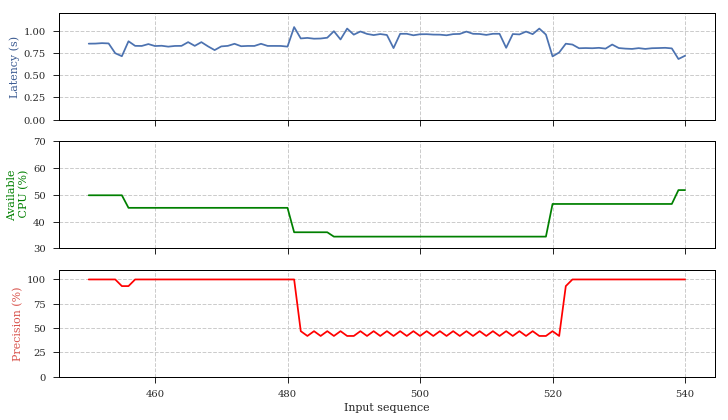}
\caption{Response to changes of RL2 based orchestration}\label{fig:rapid-change}
\vspace{-1em}
\end{figure}

Regarding the requirements for the dynamic orchestration, our RL based approach uses less than 0.5\% of the total execution time of the application, and produces even a lower overhead than the Heuristics based approach. This is visible in the results
obtained in Table~\ref{tab:adaptive-logic-requirements}. In addition, the learning
based orchestration adapts rapidly to changes as it can be seen in
Figure~\ref{fig:rapid-change}.
When the CPU availability drops around step 485, the logic automatically
changes the configuration, reducing the expected precision but managing to keep
functioning within the latency requirement. Afterwards, when the CPU
availability increases again in step 515, the orchestration changes again the
configuration to take advantage of the higher resources availability and
improving the expected precision.

\begin{table}[]
{
\setlength{\extrarowheight}{2pt}
\setlength{\tabcolsep}{5pt}
\begin{tabular}{lcccccl}
\hline
                                           &          & \multicolumn{2}{c}{Static}                                                                                                   & \multicolumn{3}{c}{Elastic} \\ \hline
                                           & Input    & \begin{tabular}[c]{@{}c@{}}High \\ Precision\end{tabular} & \begin{tabular}[c]{@{}c@{}}High Lat.\\ Satisfaction\end{tabular} & Heur.   & RL1     & RL2     \\ \hline
\multirow{4}{*}{\rotatebox[origin=c]{90}{Precision}}                 & Fixed    & 100                                                         & 41.95                                                           & 82.71  & 91.92  & 91.44  \\  
                                           & Variable & 100                                                         & 41.95                                                           & 75.81  & 76.36  & 82.66  \\  
                                           & Full day & 100                                                         & 41.95                                                           & 69.99  & 68.65  & 77.79  \\ 
                                           & Random   & 100                                                         & 41.95                                                           & 66.26  & 72.21  & 83.45  \\ \hline
\multirow{4}{*}{\rotatebox[origin=c]{90}{\parbox[c]{1.5cm}{\centering \% Latency Satisfaction}}} & Fixed    & 79.06                                                     & 95.78                                                            & 93.02   & 94.72   & 95.00   \\  
                                           & Variable & 66.38                                                     & 84.97                                                            & 82.43   & 81.36   & 79.04   \\ 
                                           & Full day & 72.15                                                     & 82.22                                                            & 79.20   & 79.31   & 79.06   \\ 
                                           & Random   & 67.27                                                     & 86.19                                                            & 80.22   & 80.89   & 83.03   \\ \hline
\end{tabular}}
\vspace{0.1cm}
\caption{Performance of static and elastic services, using different dynamic orchestration approaches}\label{tab:results}
\vspace{-3.5em}
\end{table}

\begin{table}[!htbp]
\vspace*{-\baselineskip}
\begin{tabular}{llccc}
\hline   &       & \textbf{\begin{tabular}[c]{@{}c@{}}Orchestration\end{tabular}} & \textbf{Total} & \textbf{\begin{tabular}[c]{@{}c@{}}Impact \\ in total latency\end{tabular}} \\ \hline
\multirow{2}{*}{\textbf{\begin{tabular}[c]{@{}l@{}}Intel Core \\ i5-3210M \end{tabular}}} & \textbf{Heur.} & 0.29ms            & 70.52ms       & 0.42\%   \\ & \textbf{RL}         & 0.21ms          &      70.44ms       & 0.30\%                                          \\ \hline
\multirow{2}{*}{\textbf{\begin{tabular}[c]{@{}l@{}}Raspberry \\ Pi 3 B+\end{tabular}}}                             & \textbf{Heur.} & 1.31ms*  & 313.24ms*      & 0.42\%* \\  & \textbf{RL}         & 0.93ms*     & 312.86ms*      & 0.30\%*  \\ \hline
\end{tabular}
\vspace{0.1cm}
\caption{Overhead of the orchestration to the overall service. Total execution times are calculated as the time used for the orchestration and processing a frame of 6 faces using the fastest parameter configuration. (*) projected times.}\label{tab:adaptive-logic-requirements}
\vspace{-2.5em}
\end{table}

confi
\section{Related Work}
There is extensive research in the fields related to our work, such as
self-adaptive systems and utilization of AI to optimize the performance of
applications.

Natural Adaptive Video Streaming with Pensieve~\cite{mao2017neural} presents a
system that generates adaptive bit rate (ABR) algorithms using Reinforcement
Learning. These algorithms are used for video streaming and must balance a
variety of QoE goals. This work successfully uses a variant of deep RL, A3C,
to create algorithms that adapt to a wide range of environments and QoE. The NN model runs on the server in order to avoid the overhead on the client,
something which is not always possible in edge computing. 

In Chameleon~\cite{jiang2018chameleon}, the performance of video analytics
applications is optimized by performing automatic adaptation of its
configurations. The application's behavior is customized to the execution
context by selecting different parameter configurations; the best parameter
configuration is selected by a logic inspired by greedy hill climbing combined
with periodical online profiling. However, this research is centered around
applications that use deep convolutional neural networks for video analytics,
while we aim to offer a flexible solution that can be applied to any kind of
application or service.

\section{Conclusions}
Thanks to Elastic Services, developers are able to create services that adapt
to their current execution context, achieving high requirement satisfaction
rates. By using our RL based orchestration to adapt their behavior, services
can achieve an even better performance with a lower overhead to the system.

This work has shown the efficiency of the RL based orchestration through multiple
simulations, improving on the results of the Heuristics based approach. The RL based orchestration achieves a 10--25\%
higher precision, while having a smaller system overhead, consuming 25\% less
execution time.
In addition, thanks to its capability of learning through its own experience, the
RL based orchestration offers a flexible and adaptable logic that can be used with
very different services.

In the future, we plan to optimize our approach to be able to handle a large number of parameters and parameter values, something that is challenging for the tabular Q-learning method used in this work. In order to do so, we are looking into hierarchical RL models that will be distributed between the different processing nodes.

\begin{table}
\begin{tabular}{m{0.28\linewidth}m{0.62\linewidth}}
\includegraphics[width=\linewidth]{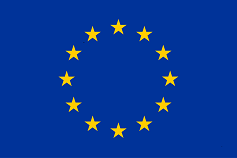} & \small{The research leading to these results has received funding from the European Community's Horizon 2020 research and innovation programme under grant agreement n\textsuperscript{o} 779747.}
\end{tabular}
\vspace{-3ex}
\end{table}

\printbibliography

\end{document}